# Anticorrelation Between Flux and Photon Index of Hard-X-Ray Emission from the Crab

Koothodil Abhijith Augustine[1] and Hsiang-Kuang Chang[1,2,3,4]
[1] Institute of Astronomy, National Tsing Hua University, Hsinchu 300044, Taiwan; abhijithaugustine007@gmail.com, hkchang@mx.nthu.edu.tw
[2] Department of Physics, National Tsing Hua University, Hsinchu 300044, Taiwan
[3] Center for Theory-Computation-Data Science Research (CTCD), National Tsing Hua University, Hsinchu 300044, Taiwan
[4] Institute of Space Engineering, National Tsing Hua University, Hsinchu 300044, Taiwan

## Abstract

Using Swift Burst Alert Telescope event-mode data during gamma-ray burst occurrences, we conduct spectral analysis for the Crab system. From 38 good observations, which span a period of 18 yr from 2006 to 2023, we find that the Crab's X-ray flux not only flickers but also significantly anticorrelates to its spectral power-law photon index. Since the emission contribution of the Crab pulsar in this energy range is small, this anticorrelation is mainly about the emission of the Crab nebula. We suggest that this anticorrelation provides observational supporting evidence for the long-standing notion that the nebula emission is due to synchrotron radiation of shocked pulsar winds in the nebula.

*Unified Astronomy Thesaurus concepts:* Millisecond pulsars (1062); Pulsar wind nebulae (2215); Pulsars (1306)

## 1. Introduction

The Crab is one of the most widely studied "stable" and bright astrophysical sources in the high-energy regime. Its central pulsar, PSR B0531+21, is a rapidly rotating neutron star that emits radiation across the electromagnetic spectrum (see J. J. Hester 2008; R. Bühler & R. Blandford 2014 for reviews on the Crab). The Crab has been used to calibrate multiple instruments in the X-ray and soft-gamma-ray regimes (M. G. Kirsch et al. 2005).

The emission from the Crab is composed of two components: from the pulsar wind nebula (PWN) and from the pulsar. The pulsar drives electromagnetic radiation through energetic flows of electron–positron pairs, which produce the pulsar component of the emission when they are still in the magnetosphere and wind zone of the pulsar. They also create a PWN shock (forward shock) and a termination shock (reverse shock) upon interacting with the interstellar medium (ISM), leading to synchrotron emission (P. Slane 2017). The synchrotron radiation from electron–positron pairs in the PWN, which manifests itself as the nebula's glow from the radio to gamma-ray bands, provides insights into particle acceleration, magnetic fields, and shock processes in the PWN (see, e.g., L. Sironi & B. Cerutti 2017; O. Arad et al. 2021). These two components, from the pulsar and from the nebula, can be distinguished through imaging or phase-resolved timing analyses, even in challenging bands like hard X-rays (e.g., M. C. Weisskopf et al. 2011; M. Vivekanand 2021). It is generally believed that in the X-ray band, the pulsar component contributes less than 20% of the total flux, with the majority coming from the PWN (L. Kuiper et al. 2001).

While the Crab pulsar is quite stable, except for a slow, secular decrease in flux (L. L. Yan et al. 2018; H.-S. Zhao et al. 2023), the nebula has been found to vary at a level of a few percent and at a seeming 3 yr timescale, with data taken by instruments on board RXTE, Swift, INTEGRAL, and Fermi in the X-ray and soft-gamma-ray bands from 1999 to 2010 (C. A. Wilson-Hodge et al. 2011) and Suzaku from 2005 to 2012 (T. Kouzu et al. 2013). What causes this variation is still not clear. It should be noted that the flux and spectral power-law photon index of the Crab based on NuSTAR measurements in 2015 and 2016 (K. K. Madsen et al. 2017) are in agreement with the values measured 42 yr ago by A. Toor & F. D. Seward (1974). This suggests that the observed variability of the Crab at timescales of years involves fluctuations around a steady mean. There is no indication of a long-term change to date.

We report in this paper our study of the Crab's variability using Swift Burst Alert Telescope (BAT) event-mode data from 2006 to 2023. Besides the flux variation at a few percent, we find this variation is negatively correlated to its spectral power-law photon index at a statistically discernible level. Such an anticorrelation may shed some light on the cause of the observed variation. We also find that the variation timescale seems to be longer than that before the year 2010.

This paper is structured as follows. First, we describe the data (Section 2.1) that are used for this study, and then how the data are analyzed (Section 2.2). We proceed to discuss the results obtained from the analysis in Section 3. Finally, we conclude the paper in Section 4.

## 2. Data

### 2.1. Data Selection

The Swift BAT is a highly sensitive, large-field-of-view (FOV), coded-aperture telescope designed to monitor a significant portion of the sky (2.2 steradians at a 10% coding fraction) for gamma-ray bursts (GRBs; S. D. Barthelmy et al. 2005). When a trigger—usually a GRB—is detected, the spacecraft autonomously slews to point its two narrow-field instruments—the X-ray Telescope (D. N. Burrows et al. 2005) and the Ultraviolet Optical Telescope (P. W. A. Roming et al. 2005)—for follow-up observations, while the BAT collects event data during this entire process. During the slew process, it is necessary to account for changes in instrumental responses during this time. Since there are ample data outside the slew period, we avoid using slew data. BAT automatically records about 1000 s of event data during onboard triggers. We utilize these event data (A. Lien et al. 2016)[5] for our analysis. We selected GRB observations by BAT that fall within a 25° FOV of the Crab nebula to maximize the $p$-code fraction. We identified 39 unique observations spanning from 2006 February to 2023 August (approximately 18 yr). We used the energy range of 15–150 keV for our analysis. While the event data record spectral information in 80 channels, which may yield more accurate spectral analysis results, the BAT survey data only record such information in eight energy bands between 14 and 195 keV. We therefore exploit the publicly available BAT 157-month survey data for the hardness ratio study only.

The BAT's localization accuracy is determined by the instrument's partial coding fraction ($p$-code)—the fraction of operational detectors exposed to a source at any given time and sky position. This $p$-code fraction is directly proportional to the instrument's sensitivity and has a maximum value of 1, indicating that all available detectors are illuminated by the desired source. The importance of the partial coding fraction lies in its impact on the counts from a source (see M. Moss et al. 2022 for more details on the $p$-code fraction).

### 2.2. Data Analysis

For data analysis, we adhered to the standard multistep data reduction process using the HEASARC (version 6.30.1) subpackage FTOOLS, as outlined in the latest (version 6.3) BAT Guide.[6] Background subtraction for BAT was achieved through a process called "mask weighting." We utilized the position of the Crab (R.A.: 83.633114, decl.: +22.01446667) for this process (Gaia Collaboration 2020), resulting in background-subtracted light curves for the Crab. During a BAT trigger, the spacecraft will often automatically slew to the trigger location. As discussed earlier, we avoid data during the slew period and instead use data from the post-slew period, which is usually longer than the pre-slew period, increasing the total source counts as the Crab moves into the BAT's FOV from an off-axis position. This rigorous process ensured that our good time interval (GTI) was uncontaminated by the corresponding GRB or burst that triggered the telescope to slew and record the event data. The partial coding fraction (refer to Table 1) of the Crab in all our data ranges from 0.58 to 1, indicating the Crab never lies close to the edge of the FOV in these observations. However, one observation (ID: 00234516000) did not yield any photons from the Crab after ray-tracing and was thus excluded, leaving a total of 38 observations. This careful selection process ensured that the data we analyzed were of the highest possible quality, free from significant contamination by other sources.

The spectrum of each observation was then constructed and fitted with a simple power-law model, using XSPEC (K. A. Arnaud 1996) version 12.13.1. Due to BAT's limited spatial resolution (a point-spread function of ∼19.′5; J. Tueller et al. 2010), the Crab is not resolved. The photon index, flux, and their chi-square values for the fitting are provided in Table 1. All errors are quoted at a 90% confidence level.

---

[5] https://swift.gsfc.nasa.gov/results/batgrbcat/
[6] https://swift.gsfc.nasa.gov/analysis/bat_swguide_v6_3.pdf

### 3. Results

The flux and photon index obtained from the spectral fitting for all the 38 observations are plotted in Figure 1. The mean flux is $1.98^{+0.04}_{-0.09} \times 10^{-8}$ ergs cm$^{-2}$ s$^{-1}$ and the mean photon index is $2.15^{+0.06}_{-0.06}$. Because of the large uncertainty, variation in the photon index cannot be confirmed statistically (fitting with a constant yielding $\chi^2_\nu = 0.58$). The flux variation is more significant with a constant fit yielding $\chi^2_\nu = 2.4$. However, they seem to vary in an opposite way. We therefore plot the flux against the photon index for all observations in Figure 2 to examine their possible anticorrelation.

In Figure 2, one can see two outliers, at the lowest and highest flux among the 38 data points, respectively. Outlier number 1, with the lowest flux, is the fifth data point from the right side in the light curve shown in Figure 1, and outlier number 2, with the highest flux, is the first data point from the left in that figure. These two outliers are from observations of ID 01104692000 (outlier 1) and ID 00180977000 (outlier 2). The Spearman correlation coefficient of this distribution, when outlier 1 is excluded, is $-0.63$, with a $p$-value of $10^{-5}$. This indicates a quite strong anticorrelation. When outlier 2 is also excluded, the Spearman correlation coefficient becomes $-0.70$, with a $p$-value of $10^{-6}$. We also fit a linear function to the 36 good points using orthogonal distance regression. The fit is marked with a gray line in Figure 2, and the $\chi^2_\nu$ for the fit is 0.24. This small $\chi^2_\nu$ is due to the large uncertainty of the 36 data points.

In view of the large uncertainty, we conducted Monte Carlo simulations to better estimate the significance of the correlation between the flux and the power-law index. We performed 100,000 simulations, each time selecting a set of 36 pairs of flux and power-law index values, from a two-dimensional Gaussian distribution based on the corresponding uncertainties. The resulting distribution of Spearman correlation coefficients (Figure 3) has a mean of $-0.34$. The standard deviation of this distribution is 0.14. This analysis indicates that, when the uncertainties are taken into account, the current data show an anticorrelation at a significance level represented by a 0.02 $p$-value (Spearman's coefficient $-0.34$, 36 data points). This is weaker than the measured one mentioned in the above paragraph but still at a discernible level.

The Crab count rate taken from the publicly available "Swift BAT 157-Month Hard X-ray Survey"[7] (A. Lien et al. 2023) is also plotted in Figure 1. This shows the variation timescale changed after the year 2010, consistent with the flux we derived from Swift BAT event data. To describe the variation timescale after 2010, we use a sinusoidal function to fit the flux and photon index, excluding outlier 1. The best-fit period for the flux is 5.92 yr ($\chi^2_\nu = 1.33$; $p$-value equal to 0.12) and that for the photon index is 5.94 yr ($\chi^2_\nu = 0.37$). The timescale is about two times longer than that before 2010 (C. A. Wilson-Hodge et al. 2011). These two best-fit curves also clearly show the anticorrelation.

An anticorrelation between the photon index and flux implies a positive correlation between the hardness ratio and count rate. The Swift BAT 157-Month Hard X-ray Survey catalog provides data recorded in eight energy bands. For the analysis, we defined the soft band as 14–50 keV and the hard band as 50–195 keV. The hardness ratio of these 157 monthly

---

[7] https://swift.gsfc.nasa.gov/results/bs157mon/287

**Table 1**
Table of GRB and Burst Observations Within 25° Radius of the Crab for a Period of 18 yr

| Observation ID | GRB or Burst | GTI (s) | p-code Fraction | Start Time (MJD) | $\chi^2$ (d.o.f = 56) | Photon Index ($\Gamma$) | Flux ($10^{-8}$ erg cm$^{-2}$ s$^{-1}$) |
|---|---|---|---|---|---|---|---|
| 00180977000 | GRB 060210 | 241 | 0.58 | 53776.20 | 56.56 | $2.16^{+0.08}_{-0.08}$ | $2.20^{+0.08}_{-0.14}$ |
| 00181126000 | GRB 060211a | 299 | 0.57 | 53777.39 | 49.34 | $2.12^{+0.09}_{-0.08}$ | $2.10^{+0.05}_{-0.13}$ |
| 00181156000 | GRB 060211b | 238 | 0.94 | 53777.65 | 39.15 | $2.14^{+0.07}_{-0.07}$ | $2.00^{+0.05}_{-0.12}$ |
| 00284856000 | GRB 070714b | 912 | 0.66 | 54027.17 | 48.32 | $2.18^{+0.05}_{-0.05}$ | $2.06^{+0.03}_{-0.07}$ |
| 00306793000 | GRB 080319d | 834 | 0.86 | 54295.20 | 60.39 | $2.13^{+0.05}_{-0.05}$ | $2.02^{+0.03}_{-0.08}$ |
| 00308812000 | GRB 080409 | 785 | 0.99 | 54544.70 | 35.92 | $2.17^{+0.05}_{-0.04}$ | $2.03^{+0.04}_{-0.07}$ |
| 00321376000 | GRB 080822b | 594 | 0.97 | 54565.05 | 38.53 | $2.17^{+0.05}_{-0.05}$ | $2.04^{+0.03}_{-0.08}$ |
| 00451191000 | GRB 110412a | 782 | 0.88 | 54700.87 | 52.76 | $2.24^{+0.06}_{-0.06}$ | $1.90^{+0.04}_{-0.08}$ |
| 00451343000 | GRB 110414a | 822 | 0.91 | 55663.30 | 43.59 | $2.20^{+0.05}_{-0.05}$ | $1.84^{+0.03}_{-0.07}$ |
| 00519211000 | Burst (42.459, 40.466) | 865 | 1.00 | 55665.31 | 42.07 | $2.17^{+0.05}_{-0.05}$ | $1.87^{+0.03}_{-0.05}$ |
| 00548927000 | GRB 130216a | 932 | 0.97 | 56020.03 | 43.19 | $2.12^{+0.05}_{-0.05}$ | $2.07^{+0.03}_{-0.08}$ |
| 00582184000 | GRB 131227A | 797 | 0.79 | 56339.92 | 38.05 | $2.13^{+0.05}_{-0.06}$ | $2.11^{+0.06}_{-0.09}$ |
| 00597722000 | GRB 140430a | 334 | 0.81 | 56653.19 | 41.35 | $2.12^{+0.08}_{-0.07}$ | $2.10^{+0.05}_{-0.12}$ |
| 00614390000 | Burst (76.734, 12.820) | 942 | 0.99 | 56777.85 | 54.36 | $2.08^{+0.05}_{-0.06}$ | $2.05^{+0.03}_{-0.06}$ |
| 00615399000 | GRB 141015a | 206 | 0.98 | 56934.96 | 56.85 | $2.11^{+0.08}_{-0.08}$ | $2.02^{+0.04}_{-0.10}$ |
| 00629578000 | GRB 150203a | 1110 | 0.97 | 56945.37 | 47.59 | $2.21^{+0.05}_{-0.04}$ | $1.94^{+0.02}_{-0.06}$ |
| 00655262000 | GRB 150911a | 1686 | 0.82 | 57056.16 | 54.64 | $2.20^{+0.06}_{-0.05}$ | $1.90^{+0.03}_{-0.06}$ |
| 00667392000 | GRB 151215a | 259 | 0.78 | 57276.77 | 44.48 | $2.22^{+0.09}_{-0.09}$ | $1.87^{+0.07}_{-0.15}$ |
| 00669319000 | GRB 160104a | 780 | 0.92 | 57371.11 | 51.86 | $2.20^{+0.05}_{-0.05}$ | $1.97^{+0.03}_{-0.08}$ |
| 00676595000 | Burst (107.316, 26.932) | 830 | 0.69 | 57391.46 | 60.9 | $2.22^{+0.06}_{-0.05}$ | $1.92^{+0.04}_{-0.07}$ |
| 00680017000 | GRB 160321A | 859 | 0.84 | 57446.72 | 60.9 | $2.22^{+0.06}_{-0.06}$ | $1.92^{+0.03}_{-0.08}$ |
| 00706052000 | Burst (98.821, −6.645) | 237 | 0.96 | 57468.65 | 41.55 | $2.23^{+0.08}_{-0.08}$ | $1.85^{+0.04}_{-0.08}$ |
| 00709765000 | Burst (80.088, 40.029) | 795 | 1.00 | 57595.05 | 65.67 | $2.17^{+0.05}_{-0.05}$ | $1.86^{+0.04}_{-0.09}$ |
| 00716127000 | GRB 161007a | 840 | 0.77 | 57624.96 | 66.49 | $2.20^{+0.05}_{-0.05}$ | $1.93^{+0.04}_{-0.06}$ |
| 00820347000 | Burst (66.024, 13.398) | 862 | 0.78 | 57668.88 | 46.2 | $2.15^{+0.05}_{-0.05}$ | $2.03^{+0.03}_{-0.08}$ |
| 00824063000 | Burst (95.957, 12.811) | 790 | 0.90 | 58208.17 | 54.34 | $2.17^{+0.05}_{-0.05}$ | $2.04^{+0.04}_{-0.07}$ |
| 00853882000 | Burst (104.225, 39.315) | 764 | 0.81 | 58218.32 | 42.88 | $2.10^{+0.06}_{-0.06}$ | $2.09^{+0.04}_{-0.08}$ |
| 00865036000 | Burst (64.746, 15.213) | 823 | 0.79 | 58348.51 | 46.27 | $2.14^{+0.06}_{-0.06}$ | $2.06^{+0.05}_{-0.09}$ |
| 00927345000 | Burst (86.368, −15.907) | 306 | 0.74 | 58393.47 | 47.5 | $2.10^{+0.08}_{-0.08}$ | $2.07^{+0.08}_{-0.14}$ |
| 00995004000 | GRB 200907b | 767 | 0.98 | 58757.71 | 41.55 | $2.16^{+0.06}_{-0.06}$ | $1.91^{+0.06}_{-0.08}$ |
| 01032183000 | Burst (72.266, 7.269) | 209 | 0.92 | 59099.77 | 49.57 | $2.10^{+0.09}_{-0.09}$ | $2.03^{+0.07}_{-0.17}$ |
| 01036227000 | GRB 210308a | 631 | 0.83 | 59257.17 | 43.22 | $2.11^{+0.07}_{-0.07}$ | $2.07^{+0.05}_{-0.09}$ |
| 01073893000 | GRB 210919a | 1305 | 0.98 | 59281.26 | 50.68 | $2.16^{+0.06}_{-0.06}$ | $2.03^{+0.03}_{-0.07}$ |
| 01090472000 | GRB 211221A | 917 | 0.64 | 59476.01 | 46.85 | $2.03^{+0.07}_{-0.07}$ | $1.66^{+0.04}_{-0.11}$ |
| 01104692000 | GRB 220430a | 725 | 0.83 | 59699.57 | 39.09 | $2.17^{+0.06}_{-0.06}$ | $1.82^{+0.04}_{-0.10}$ |
| 01104842000 | Burst (85.577, 14.033) | 541 | 0.89 | 59700.82 | 61.33 | $2.18^{+0.07}_{-0.07}$ | $1.92^{+0.04}_{-0.10}$ |
| 01125809000 | GRB 220930a | 761 | 0.79 | 59852.45 | 45.96 | $2.13^{+0.06}_{-0.06}$ | $2.08^{+0.05}_{-0.12}$ |
| 01182085000 | Burst (76.240, 19.561) | 761 | 0.99 | 60158.27 | 41.15 | $2.14^{+0.06}_{-0.05}$ | $1.97^{+0.03}_{-0.07}$ |

**Note.** The partial coding fraction (which represents the quality of the observation) and $\chi^2$ are given, for estimating the quality of the fit. The GTI is the amount of time the post-slew spectral analysis was done for. The observation with ID 00234516000 did not yield any photons while the light curve was produced. The results from the spectral analysis for the remaining 38 observations are given in this table. The photon index and flux in the energy range of 15–150 keV are also provided.

average values against their count rates is plotted in Figure 4. At first glance, the 157 points do not show a clear correlation. Noticing that the count rate has a deep drop in 2010 and the count-rate level is lower after that, we separated the 157 points into the first 72, corresponding to the first 6 yr of data from 2004 December, represented by blue circles, and the last 85, corresponding to the later 7 yr of data up to 2017 December, represented by orange boxes. A weaker positive correlation was found for the first group, with a Spearman coefficient of 0.26, corresponding to a p-value of 0.03. A much stronger one was found for the last group, with a 0.56 Spearman coefficient, which gives a highly significant p-value of $10^{-8}$. It is not clear what happened in 2010 when the Crab had a deeper drop in its count rate.

## 4. Discussion and Conclusion

Electromagnetic radiation of the Crab nebula, from radio to gigaelectronvolt (GeV) gamma-ray bands, is usually considered to be due to synchrotron radiation, because of its nonthermal spectral shape and high degree of polarization. That beyond GeV energy is attributed to coming from inverse-Compton scattering between those synchrotron radiation photons (plus cosmic microwave background) and relativistic pair plasma in the pulsar winds. The synchrotron component, however, has a photon index increasing with energy—that is, it is softer toward higher-energy bands. That, in turn, requires the radiating electrons to have corresponding power indices in their energy distribution, also increasing with energy. There are still many unsolved issues as to how to accelerate electrons to that kind of distribution (L. Sironi & B. Cerutti 2017;

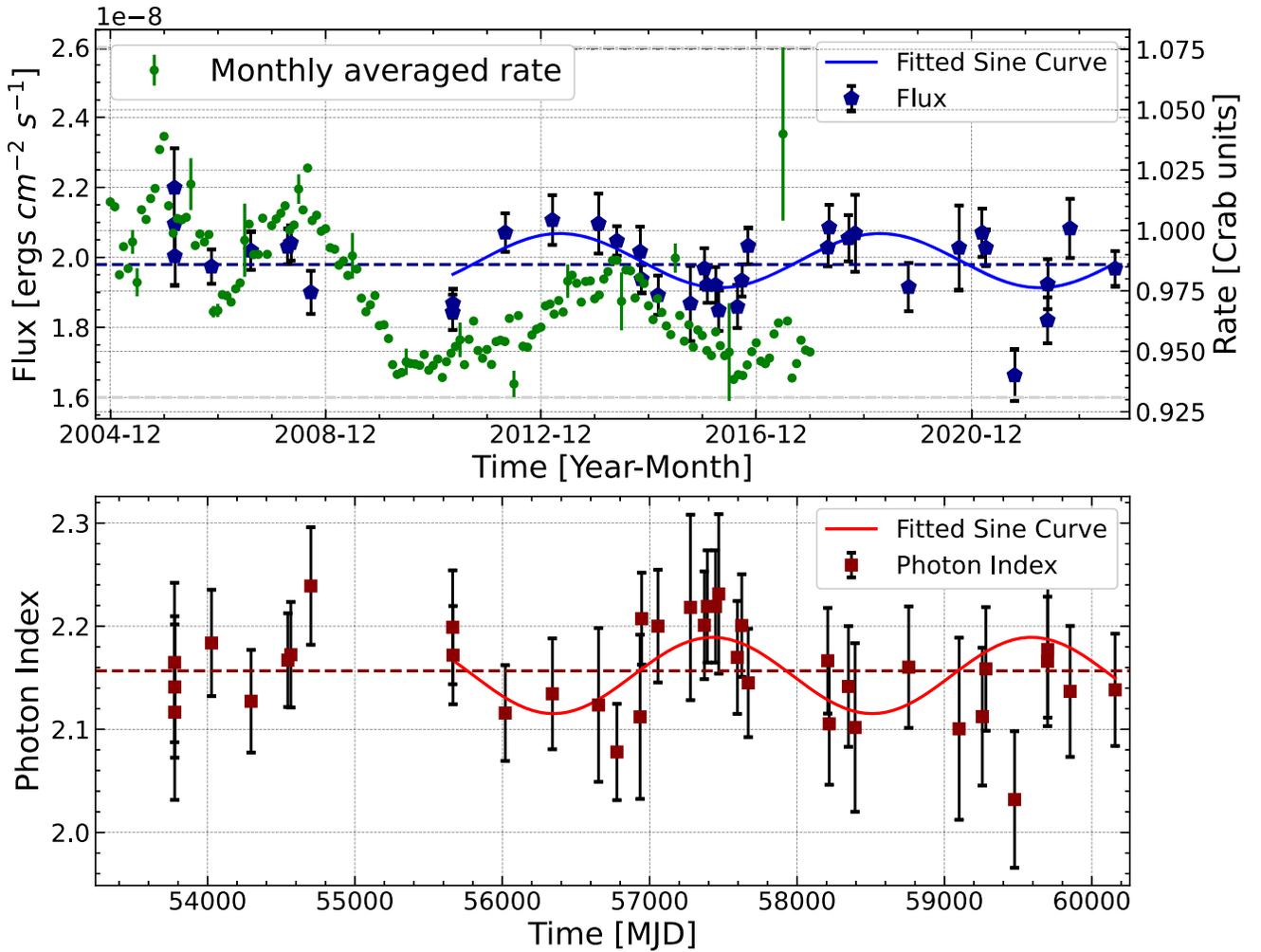

**Figure 1.** The Crab flux and photon index over 18 yr (2006 February–2023 August) measured by Swift BAT (15–150 keV). The upper panel shows the flux and the lower panel shows the photon index. Time is given in MJD and years. The dashed lines indicate the average flux and the average photon index of the 38 good observations. The green points represent the count rate in Crab units sourced from the publicly available "Swift BAT 157-Month Hard X-ray Survey" in the 14–195 keV energy band. The two sinusoidal curves are the best-fit sinusoidal models to the data after the year 2010, excluding the outlier at MJD 59476.

O. Arad et al. 2021). Nonetheless, no matter whether pulsar winds are energized on the way before reaching the termination shock by magnetic reconnection (J. Pétri 2012) or turbulent magnetic relaxation (J. Zrake & J. Arons 2017) in the striped pulsar winds, shock acceleration at the termination shock should still be at work to form nonthermal populations of relativistic electrons, which emit the observed synchrotron radiation.

The major finding of this study—that is, the anticorrelation between the hard-X-ray flux of the Crab and its spectral photon index—is likely a manifestation of synchrotron radiation, due to particles accelerated by diffusive shock acceleration (DSA) in the PWN. On one hand, the relativistic electrons of a power-law energy distribution, $N_{E,e} \propto E^{-p}$, can lead to a power-law synchrotron radiation of $F_\nu \propto \nu^{-(\frac{p-1}{2})}$—or, put in the photon number flux density format, $N_{E,\gamma} \propto E^{-(\frac{p-1}{2})-1} = E^{-(\frac{p+1}{2})} = E^{-\Gamma}$, where $\Gamma = \frac{p+1}{2}$ is the photon index. Moreover, DSA may create a power-law energy distribution of electrons described as (Equation (17.30) in M. S. Longair 2011)

$$N_{E,e} \propto E^{-\left(1-\frac{\ln P}{\ln \beta}\right)}, \quad (1)$$

where $P$ is the probability of particles remaining in the acceleration region after one collision and $\beta$ is the multiplicity for energy gain after one collision. We therefore have the power index $p$ in the energy distribution of electrons as $p = 1 - \frac{\ln P}{\ln \beta} = 1 + \frac{|\ln P|}{\ln \beta}$, noting that $P < 1$ and $\beta > 1$. On the other hand, the change in the synchrotron radiation flux may be due to a change in the radiating particle flux or in the magnetic field strength. The latter actually provides a possible link to the change of the photon index. When the field strength increases, not only does the radiation flux increase, but the remaining probability $P$ in Equation (1) also increases, because of magnetic irregularity or turbulence with increased field strength. The energy multiplicity factor $\beta$ is probably not sensitive to the magnetic field strength. With a larger $P$, the power index $p$ of the electron energy distribution is smaller, and so is the radiation photon index $\Gamma$. The synchrotron radiation flux is therefore anticorrelated with its photon index if the energetic electrons are accelerated by DSA and the change is due to magnetic field strength change.

The magnetic fields around PWN shock fronts are compressed interstellar fields in the course of the nebula expansion. These fields may have been modulated in some way by the Crab supernova ejecta, which are considered to have dispersed into the ISM. The details of this modulation are not clear, but we noticed that the X-ray flux variation shows a roughly 3 yr timescale before 2010 April, when a deeper drop in flux occurred and after which the variation timescale seems to lengthen (Figure 8 in K. Oh et al. 2018). The new timescale is about 6 yr, as can be seen in Figure 1. These timescales might be relics of earlier ejecta–ISM interactions.

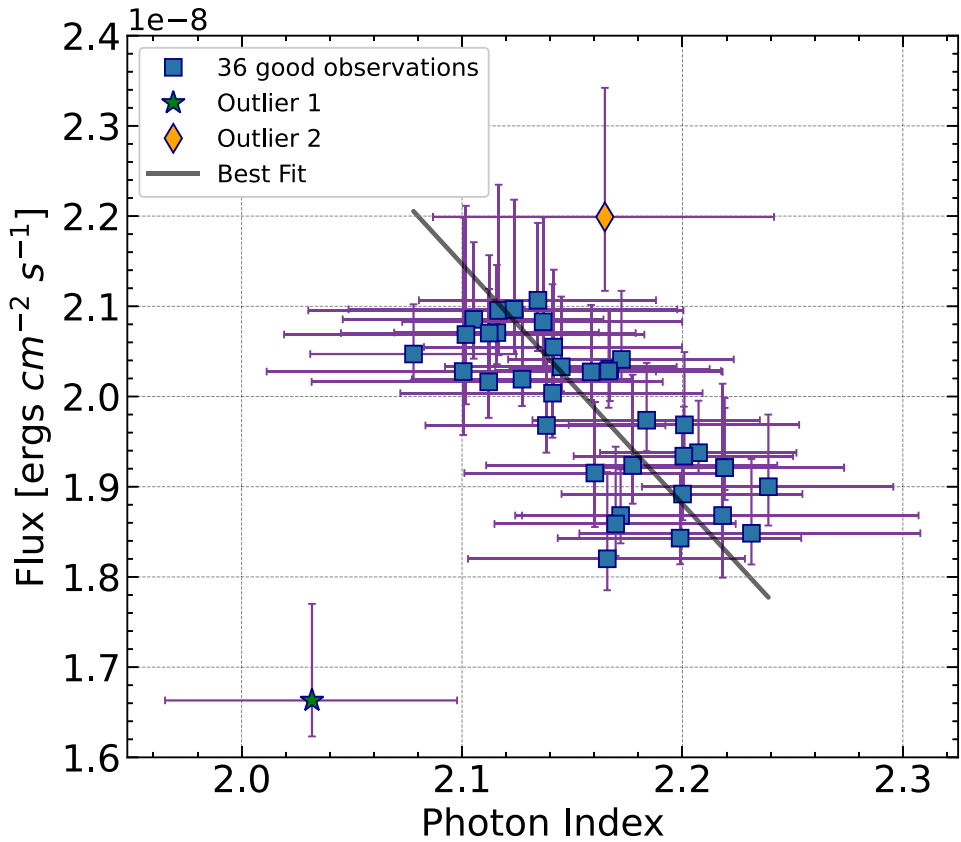

**Figure 2.** Flux vs. photon index for all the 38 BAT (15–150 keV) observations. The two outliers are marked in different colors and shapes. The black line is the best linear fit of these data points, excluding both outliers, obtained from the orthogonal distance regression method.

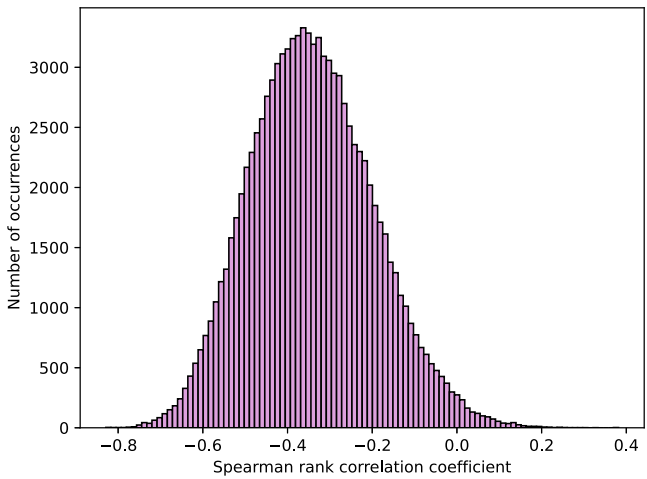

**Figure 3.** Histogram of Spearman correlation coefficients from 100,000 Monte Carlo simulations. The mean Spearman correlation coefficient is at $-0.34$.

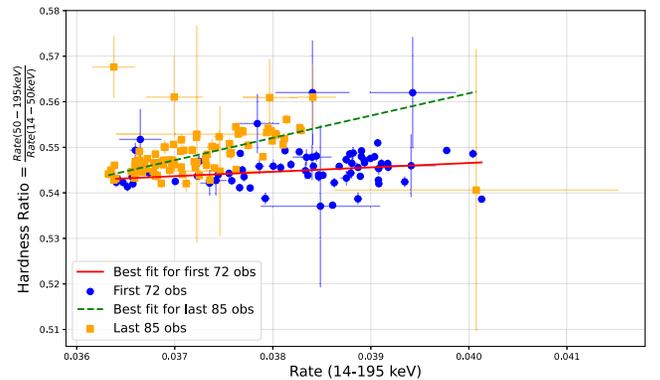

**Figure 4.** Hardness ratio plot derived from the Swift BAT 157-Month Hard X-ray Survey catalog, utilizing eight energy bands. The soft band (14–50 keV) and hard band (50–195 keV) were used to calculate the hardness ratio. Observations from the first 6 yr (2004 December onward) are depicted as blue circles, while data from the subsequent period up to 2017 December are represented as orange boxes. The Spearman's rank correlation coefficients for the first and later observations are 0.26 ($p = 0.03$) and 0.56 ($p \approx 10^{-8}$), respectively.

The anticorrelation of the X-ray flux and photon index in 15–150 keV that we report here has actually already revealed itself in the upper panel of Figure 2 in T. Kouzu et al. (2013), with a positive correlation between spectral hardness and count rates, using Suzaku data. It is also consistent with the anticorrelation that is obtained from the Swift BAT 157-Month Hard X-ray Survey data (shown in Figure 4). This anticorrelation supports the idea that the hard-X-ray emission from the Crab is synchrotron radiation of energetic electrons accelerated by DSA and the variation is due to a magnetic field strength change around the shock front. The flux and spectral variation in the emission from the Crab nebula probably do not appear only in the hard-X-ray regime. At higher energies, the Crab is not spatially resolved and the pulsar contribution to the whole (nebula-plus-pulsar) emission increases with energy (L. Kuiper et al. 2001). It is therefore not easy to examine this variation. On the other hand, in soft X-rays or lower-energy bands, the Crab can be resolved and the nebula dominates the whole emission. Similar studies in other energy bands are very much desired. In addition, the polarization degree of these emissions may also vary in a way that is positively correlated with the variation of the photon index. This may be verified in soft X-rays by IXPE in the near future (N. Bucciantini et al. 2023).

## Acknowledgments

This work is supported by the National Science and Technology Council (NSTC) of the Republic of China (Taiwan) under grant 112-2112-M-007-053. The results presented here are obtained by using data obtained by the BAT on the Swift observatory by NASA with the participation of Italy and the UK. We are very grateful to Amy Lien of the University of Tampa for her help with the data processing in this work and to the anonymous referee for the valuable comments, which improved this paper significantly.

## ORCID iDs

Koothodil Abhijith Augustine  https://orcid.org/0009-0008-7888-7584
Hsiang-Kuang Chang  https://orcid.org/0000-0002-5617-3117